Contact details corresponding author:
Max Kemman, Erasmus University Rotterdam, Rotterdam, The Netherlands.
Email: kemman@eshcc.eur.nl

# Talking with Scholars: Developing a Research Environment for Oral History Collections

Max Kemman[1], Stef Scagliola[1],
Franciska de Jong[1,2], and Roeland Ordelman[2,3]

[1] Erasmus University Rotterdam, Rotterdam, The Netherlands
{kemman, scagliola}@eshcc.eur.nl
[2] University of Twente, Enschede, The Netherlands
f.m.g.dejong@utwente.nl
[3] Netherlands Institute for Sound and Vision, Hilversum, The Netherlands
rordelman@beeldengeluid.nl

**Abstract.** Scholars are yet to make optimal use of Oral History collections. For the uptake of digital research tools in the daily working practice of researchers, practices and conventions commonly adhered to in the subfields of the humanities should be taken into account during development, in order to facilitate the uptake of digital research tools in the daily working practice of researchers. To this end, in the *Oral History Today* project a research tool for exploring Oral History collections is developed in close collaboration with scholarly researchers. This paper describes four stages of scholarly research and the first steps undertaken to incorporate requirements of these stages in a digital research environment.

## 1    Introduction

The digital turn has profoundly influenced historical culture and has led to a rise in the creation of audio-visual archives with personal narratives, commonly identified as Oral History. For the general public, searching these archives by making use of standard search tools may be sufficient. Yet for scholars, the full value of this type of data cannot be exploited optimally as available tools do not enable scholars to engage with the content for the purposes of research.

When working with audio-visual content, the availability of annotations is key to the process of digging up interesting fragments. In the past years, a lot of effort has been put in tools for creating manual annotations and generating annotations (semi-)automatically. But to accelerate scholarly research, tools are required that can take available annotation layers as input and provide means for visualization, compression and aggregation of the data. Thus allowing the researcher to explore and process the data, both at fragment-, item- and collection-level.

However, to develop such dedicated data exploration tools, technology specialists and researchers in the humanities have to engage in a process of mutual understanding and joint development. Taking carefully into account the specific set of practices and conventions commonly adhered to within the subfields in the humanities is a minimum requirement for the uptake of the technology in the daily working practice of scholars. In this paper we present a research tool developed in close collaboration with scholars that enables searching and exploration of aggregated, heterogeneous Oral History content.

## 2  Four Stages of Scholarly Research

The user interface development is based upon four stages of scholarly research that were defined on the basis of an investigation of use scenarios reported in [1].

**Exploration and selection.** In the first stage, the focus is on the exploration and selection of one or more content sets within an archive that may be suitable for addressing a certain scholarly issue. The first steps in content exploration by a researcher often come down to searching for material. Research starts with the search for new or additional data. This stage can get the form of plain browsing, but it can also be strongly purpose-driven, (e.g., checking details, searching for complementary sources), item-oriented (e.g., finding the first interview with a specific person), or directed towards patterns in a collection, in which case an entire data set is the focus of attention.

**Exploration and investigation**. Once the relevant materials have been identified, the focus in the next stage is mostly on the further exploration of the collected materials, the ordering, comparison (by individual researchers or in joint efforts) and analysis, and the documentation of the interpretation. This exploration stage may generate new ideas and perspectives, requiring new searches and inquiries.

**Result presentation.** After the analysis has been completed, the third stage is the presentation of research results. In the digital realm it has become feasible to link annotations that capture the results of an analytical step to the data on which they are based. Data and annotations can be shared with peers, both during collaboration as well as in publications. Instead of a printed book, one can produce a digital publication with links to audio-visual content.

**Data curation.** The fourth and final stage of the process is the long-term preservation of the data and the results of the investigation that has been carried out. Especially audio-visual materials that have been processed with digital tools are not the kind of research result that can be stored in a cupboard; they should be deposited in a trusted digital repository [2]. Ideally the depositing of material should be in line with emerging standards for Open Data, as this would allow the data and annotations to be reused by scholars with similar interests. For example, links can then be created to other data sets to place the data in a broader context [3]. Although the actual curation process itself is out-of-scope in this specific research project, workspaces can provide a form of data curation through the individual collecting of interviews, cutting interesting fragments with a virtual cutter and creating additional manual annotations that can be fed into the existing metadata and thereby enrich the collection even further.

## 3   Oral History Today Research Environment

**Visual search.** The *Oral History Today* research interface is based upon the four stages described above. As the search process for the *exploration and selection* and *exploration and investigation* stages is reminiscent of Shneiderman's *Visual Information-Seeking Mantra* of *overview first, zoom and filter, then details-on-demand* [5], we developed a visual search interface to provide overview and zooming facilities, as well as support exploration strategies.

Two visualizations were developed to complement the search interface and allow visual searching: *word clouds* and *visual facets*. Word clouds provide a textual insight in the material available, with the additional benefit that a better insight is gained in what terminology is used in the collections explored; an issue identified for keyword search interfaces [4]. Visual facets (Fig. 1) provide a visual overview of the facets. Facets are shown as graphical bars, where the length of each value represents the number of related search results, as demonstrated previously in Relation Browser++ [6]. A difference with RB++ is that the facet values are stacked into a single bar representing the facet. On mouse-hovering a tooltip is shown with a textual description and the number of corresponding items. When the user selects a facet value, the facet bar is moved to the top to allow the user to keep a history of selected facets. Visual facets not only give a more visual overview of the search results, but also allow for faster interactions with the facets.

**Evaluation.** To allow user feedback to be incorporated in the development process, evaluation is undertaken in multiple cycles. To elicit a broad range of responses with regard to usability as well as applicability to research practices, the first cycle was performed with semi-structured interviews. Five scholars were asked to try research subjects of their own interest. The results of this first evaluation are very positive. Concerning the visualisations described above, it was generally agreed that word clouds enable the searcher to acquire an idea of what material is available. However, they did not think word clouds would provide them with keywords to improve their queries. Visual facets were considered interesting and felt as a very fast way to both acquire an overview of the search results as well as refine search results.

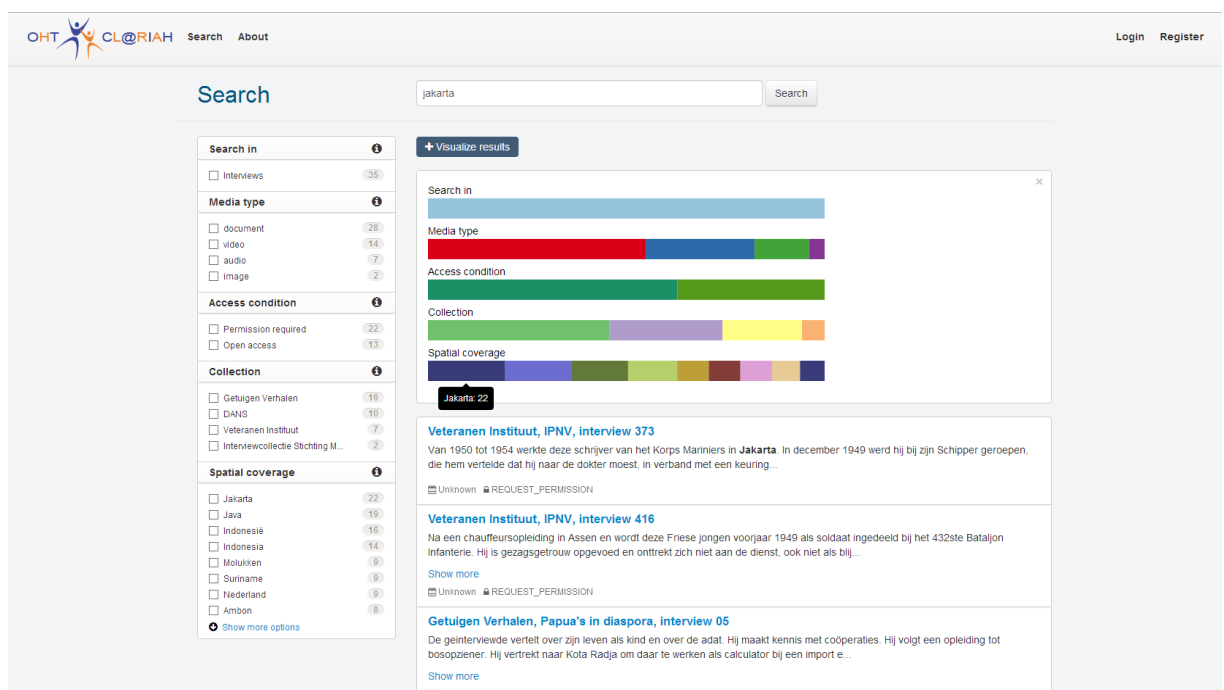

**Fig 1.** Visual Facets

**Further adjustments.** Scholars noted that being able to quickly assess the importance of search results is vital during the *exploration and selection* stage. To enable fast assessments, we added the ability to expand summary-descriptions in the search results, no longer requiring scholars to open each individual search result. After this assessment, scholars need to be able to save important items. Therefore, we developed workspaces, which allow researchers to save interviews in project-specific sets for later analysis, as well as for referencing in publications as needed in the *result presentation* stage described above.

## 4     Conclusion

The results of the first evaluation are promising. The positive responses of the scholars indicated that the chosen approach for exploring Oral History data is in the right direction. In the near future, this evaluation will receive a larger follow-up in the final evaluation of the research interface. After this final evaluation, the tool will be released to the Oral History research community, allowing us to investigate how it will eventually be used in daily research practices.

**Acknowledgements.** The work reported in this paper was funded by the EU Project AXES - *Access to Audiovisual Archives* (FP7-269980) and the Dutch national program CLARIAH (http://www.clariah.nl/). We thank Dispectu (www.dispectu.com) and Spinque (www.spinque.nl) for their collaboration in the research project *Oral History Today*.